\documentclass{aa}
\include{psfig}
\begin{document}

   \thesaurus{08     
              (08.06.3;  
               08.12.3;  
               08.08.1;  
               10.19.2)}   

\title{Fundamental  parameters of   nearby stars 
from the comparison with evolutionary calculations: masses, radii and effective temperatures}

   \author{C. Allende Prieto
          \and
          D. L. Lambert
         }

   \offprints{C. Allende Prieto}

   \institute{McDonald Observatory and Department of Astronomy, The
		University of Texas at Austin,
              RLM 15.308, Austin, TX 78712-1083\\
              email: callende@astro.as.utexas.edu, dll@astro.as.utexas.edu
             }

   \date{Received ; accepted }
   \authorrunning{Allende Prieto \& Lambert}
   \titlerunning{Fundamental parameters of stars}

   \maketitle

   \begin{abstract}

The {\it Hipparcos} mission has made it possible to constrain the
positions of nearby field stars in the colour-magnitude diagram with
very high accuracy.  These positions can be compared with the
predictions of stellar evolutionary calculations to provide information
on the basic parameters of the stars: masses, radii, effective
temperatures, ages,  and chemical composition.
  The degeneracy between  mass, age, and metallicity is not so large as
to prevent a reliable estimate of masses, radii and effective
temperatures, at least for stars of  solar metallicity. The evolutionary
models of Bertelli et al. (1994) predict those parameters finely, and
furthermore, the applied transformation from the theoretical ($\log g-$
T$_{\rm eff}$) to the observational ($M_v$--B--V) plane is precise
enough to  derive radii with an uncertainty of $\sim 6$\%, masses
within 8\%, and $T_{\rm eff}$s within $\sim 2$\% for a certain range of
the stellar parameters.  This is demonstrated  by means of comparison
with the measurements in eclipsing binaries and the InfraRed Flux
Method.

The application of the interpolation procedure  in the theoretical
 isochrones to the stars within 100 pc from the Sun
observed with {\it Hipparcos} provides estimates for 17,219 stars\footnote{Table 1 is only available in electronic form
at the CDS via anonymous ftp to cdsarc.u-strasbg.fr (130.79.128.5)
or via http://cdsweb.u-strasbg.fr/Abstract.html}.

      \keywords{Stars: fundamental parameters --
                Stars: luminosity function, mass function --
                Stars: Hertzsprung--Russel (HR) diagram --
		Galaxy: stellar content
	}
   \end{abstract}

%

\section{Introduction}

Stellar evolution calculations for single stars model their evolution
 in terms of the variations of their fundamental parameters R
and $T_{\rm eff}$, as a function of time since the starting of hydrogen fusion
reactions until their final extinction. Mass is the key parameter
that decides the stars' evolution, chemical composition and other factors
 play a secondary role. The  evolution of the stars can be
plotted as tracks  in the HR diagram, some of its areas being more
crowded by paths corresponding to stars of different masses. Other
regions are completely empty, forbidden spaces which no star is supposed 
to cross.

At least in principle,  after  proper conversion from the 
theoretical to the observational plane, it is possible
 to associate the position of a star in the HR diagram with a given stellar
mass and time since its birth, or with a range of masses and times. 
The only requirements are
an accurate knowledge of the  distance from the observer to the star, and 
a pair of photometric measurements. This is a well-known method that 
has been largely applied to 
simplified cases where some constraints on  age and distance exist, 
such as well detached binaries or stellar clusters. It has also been used 
in the search for age-metallicity relationships in the Galactic Disc or the
Halo.  Most of the stellar parameters at play in the calculations are the very
same that  define the atmospheric properties, and so 
shape  the  features observed in the stellar spectra.  These
quantities have been often  estimated directly from the
spectra, and only in a few situations, commonly  for the lack of 
empirical alternatives,  has consideration been given to the
evolutionary models.

Before trying to apply the method, several important 
 questions need to be posed, such as  
how crowded is the HR diagram, or in other
words how severe is the degeneracy between age,  mass and metallicity
for a given position in the HR diagram. It is of  relevance to
demonstrate that the translation from the theoretical parameters to the
observational plane is properly done, otherwise no matter how realistic
 the calculations are, there is no hope to get useful results.
Finally,  an extensive assessment of the adequacy of the
evolutionary calculations  is required.  
All three issues can be simultaneously
answered in  applying the method to several favourable cases.
Stellar masses and radii are known with extremely high accuracy for a
bunch of  nearby eclipsing binaries (Popper 1980; Andersen 1991). 
Some of these systems have been already employed
by Schr\"oder, Pols, \& Eggleton (1997)
 and Pols et al. (1997) to  test critically their evolutionary
 calculations (Pols et al. 1995) and tune  parameters in their scheme
that take  account of convection.  Stellar radii are alternatively and
independently measured  by interferometric techniques (e.g. Richichi et
al. 1998) and also by the so-called InfraRed Flux Method
(Blackwell \& Lynas-Gray 1994). The later provides 
probably the most direct and model-independent estimate of the stellar
effective temperature. Other parameters involved are the chemical composition and the age. Although it is possible to check the metallicity estimates from
the isochrones with the results from spectroscopic measurements, and the
stellar ages can be compared with other methods, such as activity indicators 
(see, e.g., Rocha-Pinto \& Maciel 1998), or the abundances of radioactive
nuclei (see, e.g., Goriely \& Clerbaux 1999), we shall not concentrate on them 
here.

\begin{figure}
\hspace{0cm}\psfig{figure=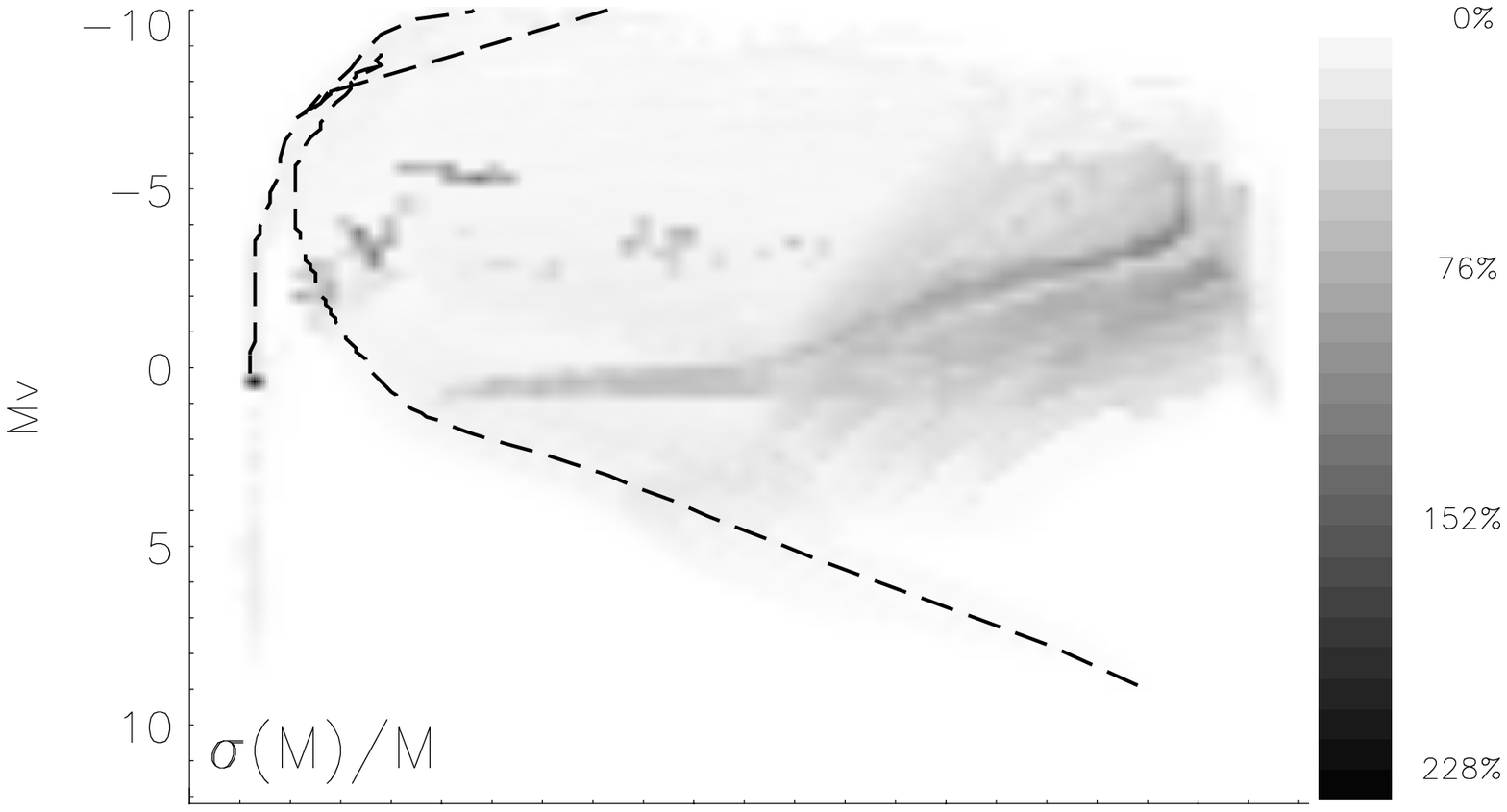,width=8.5cm}
\hspace{0cm}\psfig{figure=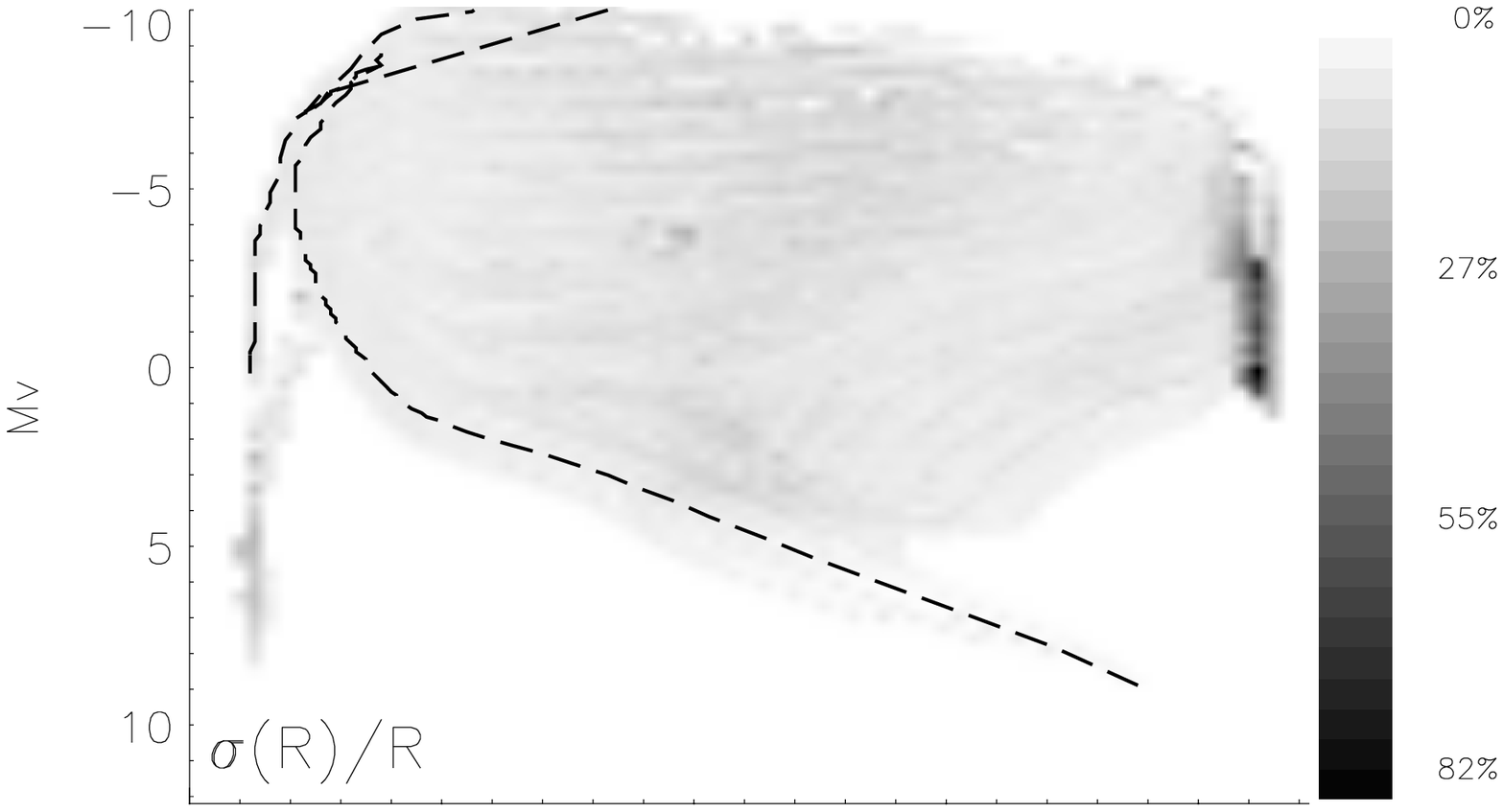,width=8.5cm}
\hspace{0cm}\psfig{figure=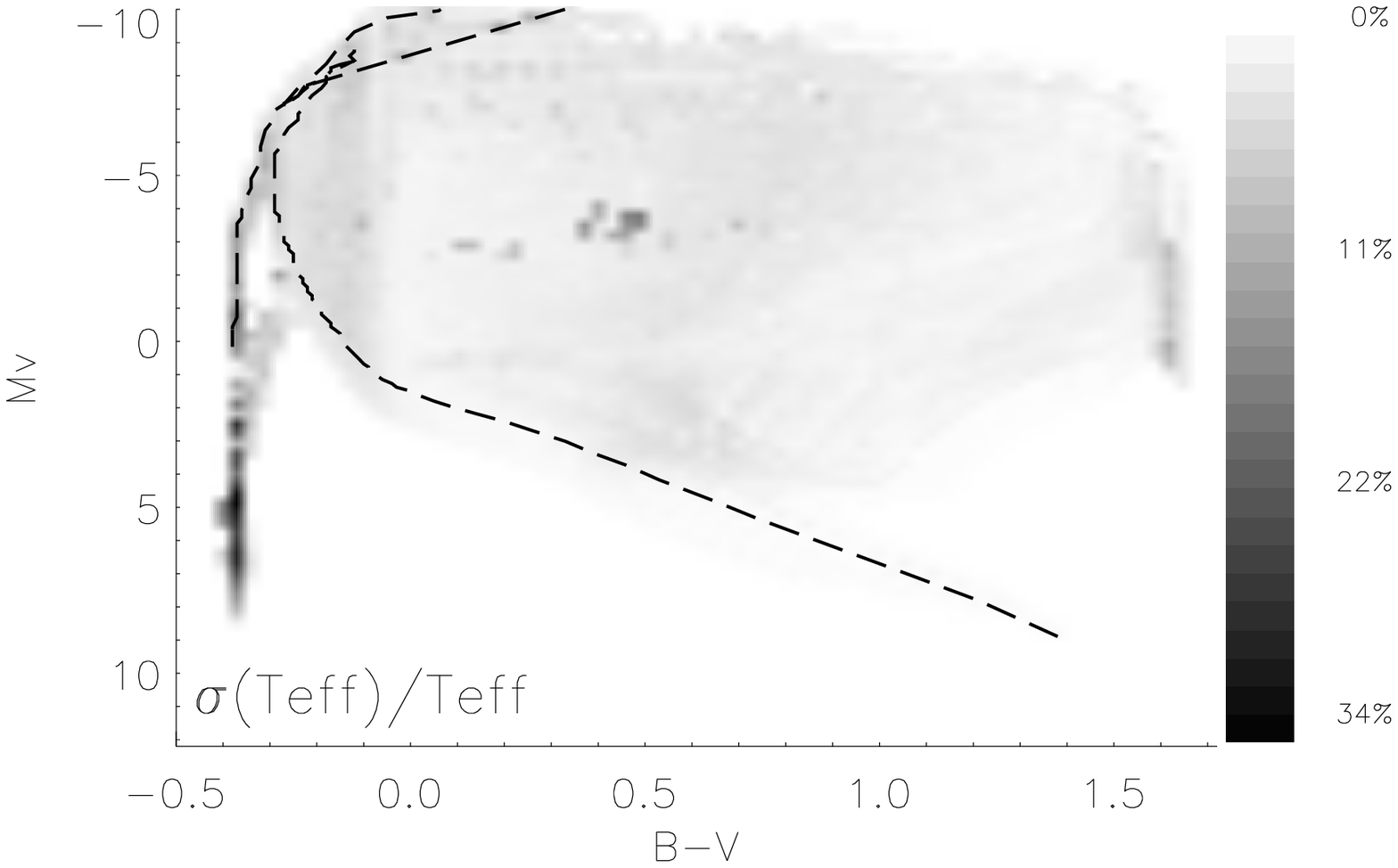,width=8.5cm}
\caption[]{Uncertainties in the colour-magnitude diagram indicating
the confusion for the masses, radii and effective temperatures retrieved for
a given star based only on its position in the diagram. The dashed line represents an isochrone of solar metallicity and 4 Myrs.}
              \label{Fig1}
 \end{figure}

In this paper, we have made use of the stellar evolutionary
calculations of Bertelli et al. (1994) to study the possibility of
retrieving the fundamental stellar parameters: radius, mass, and
effective temperature, from the comparison of the stars' position in
the colour-magnitude diagram with computations of stellar evolution for
isolated stars. We  quantify the 
 degeneracy between mass and age for a
given place in the HR diagram using error bars in the
estimates of these parameters. Stars
 in eclipsing binary systems and the InfraRed Flux Method are used to
test the procedure. And finally,  the technique is applied to  17,219 stars
identified by the {\it Hipparcos} astrometric mission 
within an sphere with  radius of 100 pc  centred in the Sun.


\section{Retrieval of the stellar parameters. Interpolation in the theoretical
isochrones}

Calculations of stellar evolution interpolated to find the position of
coeval stars of a given chemical composition but
 different masses are commonly referred to as isochrones.  Among the
published isochrones, those presented by
 Bertelli et al. (1994) span all observed stellar masses, metallicities
 from [M/H] = $-1.7$ to 0.4, and stellar ages from $ 4 \times 10^6$ to
$2 \times 10^{10}$ yrs.  The computations mostly use the OPAL radiative
opacities by Rogers \& Iglesias (1992) and Iglesias et al.  (1992).
The reader is referred to Bertelli et al. (1994) and references therein
for details.

 We assume that both the B$-$V colour, and the absolute magnitude (and
therefore the distance to the star) are
 known with enough precision to avoid significant observational errors
 affecting  derived quantities.  We quantify this statement below.
 To  retrieve the appropriate set of stellar parameters that
reproduces the position of a star in the colour-magnitude diagram we
proceed as follows. First we search the entire set of isochrones  to
find  what are, if any,  the B$-$V colours
 that reproduce the observed absolute V magnitude ($M_v$) within the
observational errors.  Then, the different possible solutions,
corresponding to different ages, metallicities, masses, etc. are
averaged  to obtain a mean value for every stellar parameter, and an
estimate of the uncertainty from the standard deviation.

The final uncertainty includes three components. First of
all, the intrinsic spread, as different evolutionary paths cross the
same area of the colour-magnitude diagram. Second, the noise introduced
in the translation from the theoretical $\log g - T_{\rm eff}$ plane to
the observational colour-magnitude diagram.  Also, observational errors
in the two considered parameters (B--V and $M_v$) contribute. Applying
the procedure along a grid of colours and absolute magnitudes, we
construct maps of the combined uncertainties expected for the different
stellar parameters.  We are mainly interested in radii, masses, and
$T_{\rm eff}$s: the magnitudes available from observations of eclipsing
binaries and the IRFM
 we are comparing to in the subsequent sections. The gray scale in the
 images included in Fig. \ref{Fig1} has been set to represent the
relative uncertainties in  these magnitudes. To construct the maps we
have assumed an uncertainty of 0.01 dex in B--V and of 0.2 dex in
$M_v$.  The dashed line indicates an isochrone of 4 Gyrs. and
 solar chemical composition.  The largest errors correspond to the
darkest areas:  $\sim$ 200 \% for the mass, $\sim$ 80\% for the radii,
and $\sim 30 \%$ for the effective temperatures.  Uncertainties in the
retrieved masses are especially significant close to the position of
the horizontal branch, where intermediate mass stars cross back from the
giant stage and more massive stars make their way up from the main
sequence, and reach the worst expectations for giants and AGB stars.
Understandably, there is also significant confusion in the area where the
post-AGB stars cross the upper main sequence way down to the white
dwarf cooling sequence, although in practice very few stars will appear
in such a rapidly evolving phase. Errors in the retrieved radii are small for
most of cases, although for the reddest evolved stars the confusion is
very large, and a similar conclusion is sketched from Fig. \ref{Fig1}
for the effective temperatures. 

These results
 show that  radii and $T_{\rm eff}$s, are  largely constrained by the
 position  of a star in the colour-magnitude diagram, regardless of the
existence of a wide range of  possibilities for  ages and
metallicities. It is apparent from Fig. 1 that the masses of evolved
stars with different ages and metallicities show more disparate values
at a given position in the colour-magnitude diagram.


\section{Critical testing}

Highly reliable measurements of stellar masses and radii are available
for eclipsing binary systems. We use them to check the 
 method suggested here for deriving the fundamental stellar parameters from
the comparison between the position of the stars in the colour-magnitude
diagram and the predictions of stellar evolution calculations. The 
model-independent effective temperatures and stellar diameters obtained
through the InfraRed Flux Method by Blackwell \& Lynas-Gray (1994)
for solar-metallicity stars are also used in the test.

\subsection{Eclipsing binary systems}

No doubt these privileged binary systems can provide the most 
solid determinations of stellar masses  and radii. 
Andersen (1991) has produced the most recent 
compilation of high-accuracy ($<$ 2\%)  determinations in binaries, listing
90 stars. Andersen lists errors for the absolute V magnitudes, and we have
assumed errors in the B--V colour  to be $0.01$ dex.

\begin{figure}
      \vspace{0cm}
	\hspace{0cm}\psfig{figure=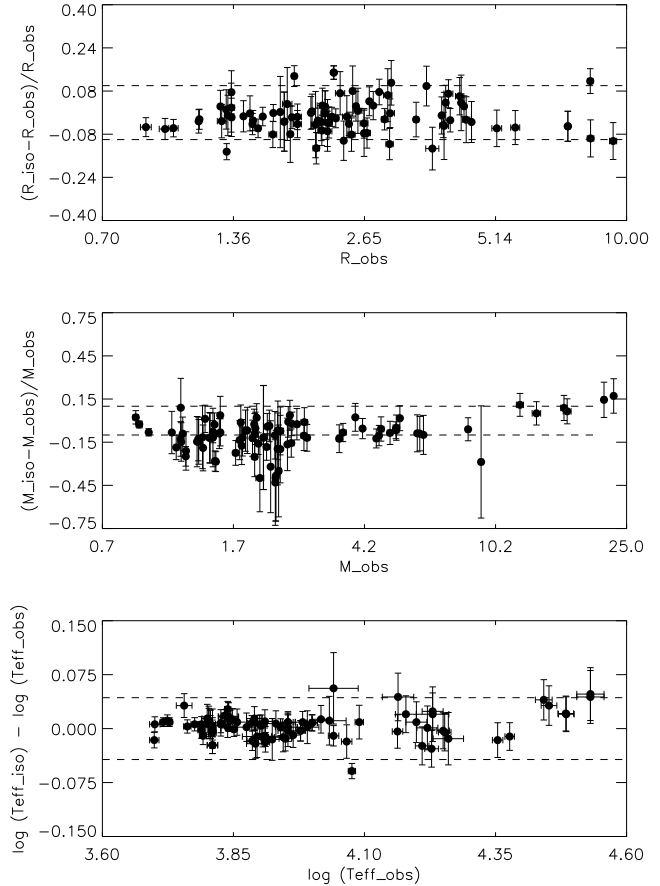,width=8.5cm}
      \caption[]{Relative differences between stellar radii, masses and $T_{\rm eff}$ derived from stellar evolutionary calculations and the robust direct
estimates from observations in eclipsing spectroscopic binaries compiled by Andersen (1991).
	}
      \label{Fig2}
      \vspace{0cm}
\end{figure}

\begin{figure}
      \vspace{0cm}
	\hspace{0cm}\psfig{figure=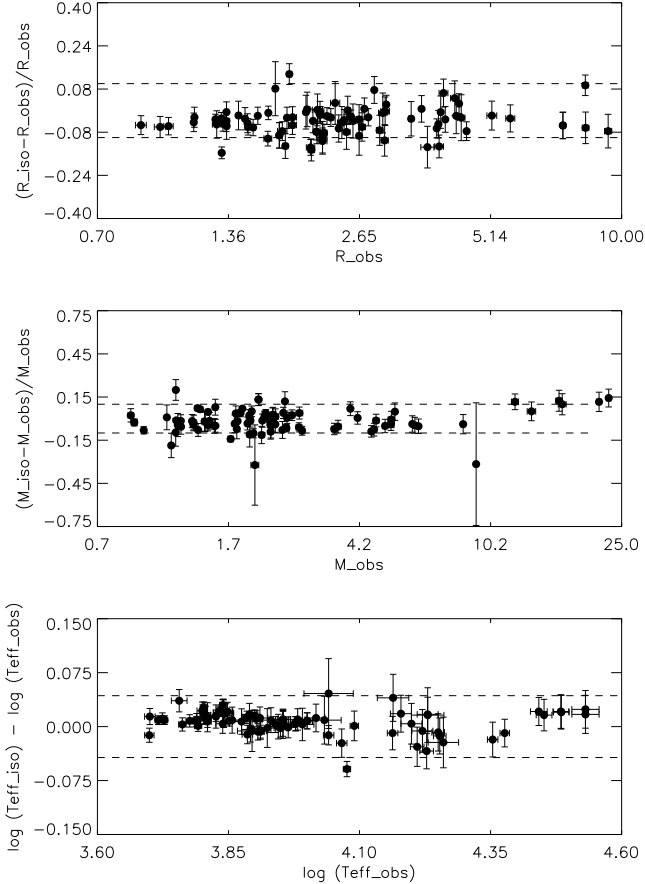,width=8.5cm}
      \caption[]{Relative differences between stellar radii, masses and $T_{\rm eff}$ derived from stellar evolutionary calculations assuming roughly solar
metallicity ($-0.4 <$ [Fe/H] $< +0.4$) and the  direct
estimates from observations in eclipsing spectroscopic binaries compiled by Andersen (1991).
	}
      \label{Fig2solar}
      \vspace{0cm}
\end{figure}

The upper and middle panels of Figure \ref{Fig2}  compare the
differences between the direct estimates of radius, and mass with those
retrieved from the interpolation in the  isochrones. The
dashed lines correspond to differences of $\pm$ 10\%.  The radii are
accurately well predicted with a remarkable standard deviation of  6\%.
The  masses  show a larger scatter of $\sim 12$\%.  Note that in these
two panels of Fig. \ref{Fig2}, the horizontal axes  have  been
logarithmically expanded for clarity.  The effective temperatures
listed by Andersen (1991)
 come from different estimators. However, it is important to remark
that spectroscopic analyses of these systems normally
 combine photometry and spectroscopy, and then the effective
temperatures are likely to be more reliable than those in most of
studies of isolated field stars. The mean of the relative errors in the
effective temperature provided by Andersen for his
 sample is 3\%. The lower panel of Fig. \ref{Fig2} shows the comparison
with the effective temperatures retrieved from the evolutionary models.
The  agreement is excellent: the  standard deviation is a mere 4\%.

For a given position in the colour magnitude diagram,  radius and
luminosity (and therefore, effective temperature) depend very weakly on
metallicity.  However, that does not apply to mass and it is reflected
in Fig. \ref{Fig1} through the large uncertainties for the retrieved
masses.  The fact that the stars in the Andersen sample are all nearby,
and therefore are roughly of solar metallicity is very useful to demonstrate
this feature.  Fig.  \ref{Fig2solar}, represents a comparison of the
same magnitudes as in Fig. \ref{Fig2}, but now the masses, radii, and
$T_{\rm eff}$s, have been estimated assuming roughly solar metallicity
($-0.7 <$ [Fe/H] $< +0.4$), and then further restricting the set of
isochrones. The agreement between observed and predicted radii improves
very little, as reflects the standard deviation of 5\%, and no
significant improvement is achieved in the effective temperature, but
the tendency to underestimate some of the masses  no longer exists:
the standard deviation of the relative differences between predicted
and observed masses is now reduced to 8\%.  These results indicate that
the combination of these masses and radii will lead to gravities
estimates with a standard deviation in $\log g$ of 0.06 dex. For
low-gravity stars of solar metallicity, a smaller mass will be derived
by assuming the star to be metal-poor, and then the average of all
metallicities underestimate  the true value, as shown in Fig. \ref{Fig2}.



Only one of the stars in Andersen's sample (TZ For A) 
has a gravity lower than $\log g = 3$
and hence the sample is restricted to objects on, or close to the main
sequence. To avoid this restriction we have extended the sample including
a few of other systems with somewhat poorer determinations. We have included the 5 resolved 
spectroscopic binaries  compiled by Popper (1980): HD16739, HD168614, $\delta$ Equ, Capella, and Spica; and 
2 detached subgiant eclipsing systems: TY Pyx and Z Her, 
also  included   in the compilation by Popper. 
The sample was completed with
the studies of $\zeta$ Aur by Bennett et al. (1996) and $\gamma$ Per by 
Popper \& McAlister (1987). The B--V for the two components of $\gamma$ 
Per have been estimated  from  their spectral type and 
the tables of Aller et al.  (1982).

\begin{figure}
      \vspace{0cm}
	\hspace{0cm}\psfig{figure=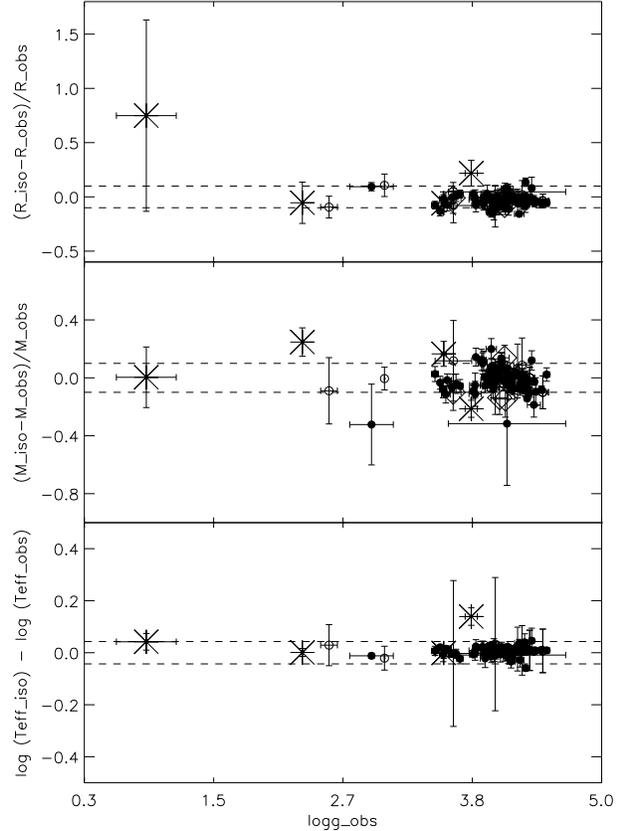,width=8.0cm}
      \caption[]{Relative differences between stellar radii, masses and $T_{\rm eff}$ derived from stellar evolutionary calculations and the robust direct
estimates from observations in eclipsing spectroscopic binaries, against
the logarithm of the  gravity. The stars compiled by Andersen (1991)  have been plotted with filled circles, the two
detached subgiant eclipsing systems (Popper 1980) 
with rhombs, the
resolved spectroscopic binaries with open circles (Popper 1980), and the two
components of $\gamma$ Per (Popper \& McAlister 1987) 
and  $\zeta$ Aur (Bennett et al. 1996) with asterisks.
	}
      \label{Fig3}
      \vspace{0cm}
\end{figure}

$T_{\rm eff}$s are provided for the stars in these systems, the
resolved spectroscopic binaries with open circles, and the two
components of $\gamma$ Per and  $\zeta$ Aur with asterisks.   Fig.
\ref{Fig3} shows no correlation of the relative differences between
{\it observed} and {\it evolutionary} radii,  masses, or effective
temperatures against $\log g$.

\subsection{The IRFM}

The InfraRed Flux Method, developed by Blackwell and his collaborators,
provides an accurate procedure to derive stellar angular diameters and
effective temperatures by measuring the monochromatic flux at an
infrared frequency and the bolometric flux, and using theoretical model
atmospheres to estimate the  monochromatic flux at the star's surface.
The method has been applied by Blackwell \& Lynas-Gray (1994) using the
Kurucz (1992) LTE model atmospheres to a sample of 80 stars. All but
one of the stars in the sample  have
 been observed by the {\it Hipparcos} mission and  we have converted
 the IRFM angular diameters to radii using the trigonometric
parallaxes ($\pi$). We have identified the position of the stars in the
colour-magnitude diagram making use of the V band photometry and the
B$-$V colours included in the {\it Hipparcos} catalogue 
 and then interpolated in
the Bertelli et al. (1994) isochrones to find  the evolutionary
status, and the fundamental stellar parameters. One star (HR1325) 
was rejected, as we could not find an isochrone close enough 
(within observational uncertainties) to the position of the star in the
colour-magnitude diagram. In this analysis,  we
have neglected the effects of reddening and assumed that the rotational
velocities are not high enough to disturb the position of the stars in
the colour-magnitude diagram. No assumption has been made about the
metallicity of the sample.

\begin{figure}
      \vspace{0cm}
	\hspace{0cm}\psfig{figure=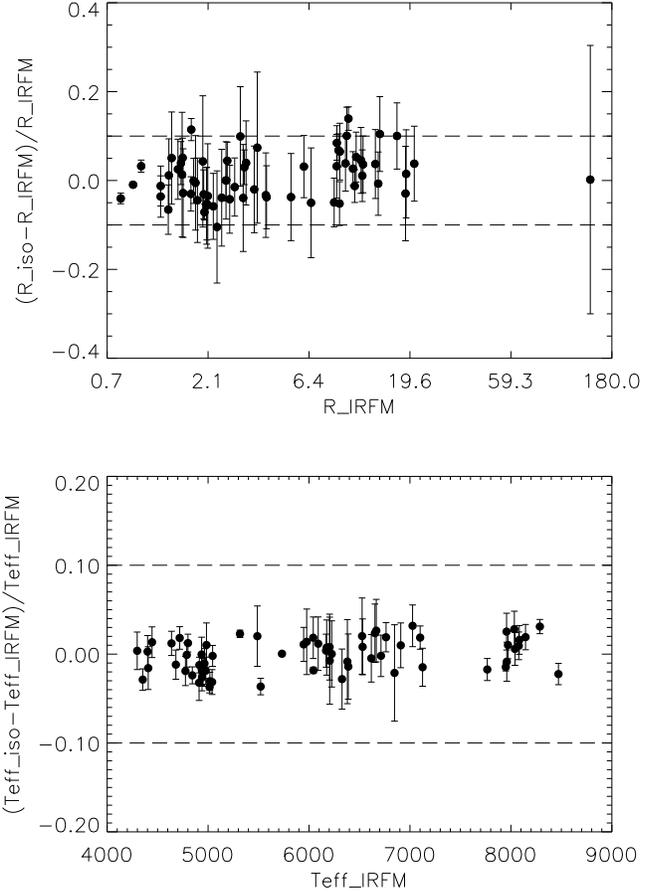,width=8.5cm}
      \caption[]{The relative differences between the radii and $T_{\rm eff}$s
 derived from stellar evolutionary calculations and those derived from the
InfraRed Flux Method (Blackwell \& Lynas-Gray 1994).
	}
      \label{Fig5}
      \vspace{0cm}
\end{figure}

The angular diameters of Blackwell \& Lynas-Gray have been converted to
radii using the Hipparcos trigonometric parallaxes.  The upper panel of
Figure \ref{Fig5} shows the comparison between the derived radii against the
estimates from the evolutionary models. Again, the agreement is
gratifying showing a standard deviation of only 5\%. The effective
temperatures retrieved from the  isochrones are also
compared to the IRFM temperatures in the lower panel of Fig. \ref{Fig5}. The
standard deviation of the comparison is less than 2\%. The sample includes
 a star as far as 277 pc from the Sun, a distance at which
 errors in the {\it Hipparcos} parallax are expected to be significant, but
the rest of the stars are  within 100 pc.

The radius of the largest star in the sample (HR2473; G8Ib) from the
IRFM measurements is about 142.0 R$_{\odot}$, in good agreement with
the prediction  from its position in the colour-magnitude diagram
142.3 R$_{\odot}$ (see Fig. \ref{Fig5}).  However,   such a highly
evolved star, as we previously found for the cool component of $\zeta$
Aur, occupies a region in the red part of the colour-magnitude diagram
quite crowded by stellar  evolutionary tracks, ending up with a very
large uncertainty in the  retrieved stellar parameters (see Fig. \ref{Fig1}). 
In the case of the radius, the standard deviation is 43 R$_{\odot}$. 
The difference here from the previous comparison for the K supergiant 
in $\zeta$ Aur is that for HR2473, the mean value of the radii 
compatible with the star's position in the colour-magnitude diagram
 is in very good agreement with the observational estimate.

\section{Application of the method to  the Hipparcos catalogue}

 Taking into account both the accuracy of the estimates and the
the range of parameters where the predictions are too vague ({\it crowded}
zones in the colour-magnitude diagram), the stellar evolutionary calculations
can be used with no more than photometry and the trigonometric parallax
within the following limits:

\begin{equation}
\begin{array}{ccccc}
0.87 &\le & R/R_{\odot} &  \le& 21 \\
0.88 &\le & M/M_{\odot} & \le& 22.9 \\
3,961 & \le&  T_{\rm eff} & \le& 33,884  ~~ {\rm K} \\
2.52 &\le&  \log g & \le &4.47. 
\end{array}
\end{equation}

The samples used here to test the accuracy of the  procedure to
retrieve the stellar parameters are strongly biased towards solar
metallicities.  We have shown that  the comparison between {\it
physical} and  predicted values is excellent when  restricting the
possible range for the metallicity to roughly solar ($-0.4 <$ [Fe/H] $< +0.4$),
 based on   {\it a priori} knowledge of the statistical peculiarities
of the sample. Nevertheless, the results here described cannot be
safely extended to low-metallicity stars without further study.

A precise knowledge of the fundamental stellar parameters is essential
to make a comparison possible  between observations and theoretical
studies, shedding light on multiple aspects of stellar structure,
stellar evolution, and the physics of stellar atmospheres.  From the
previous sections, we have seen that it is possible to use in a direct
manner the  isochrones of Bertelli et al. (1994) to
estimate masses, radii, and effective temperatures for unevolved stars
with solar-metallicity provided
 an accurate estimate of the distance from Earth is available.  The
{\it Hipparcos} mission has measured, among other quantities,
 parallaxes that lead  to distances precise typically better than 20\%
up to 100 pc from the Sun (see, e.g., Perryman et al. 1995). We have
derived absolute magnitudes and combine them with the B--V index
(compiled also in the {\it Hipparcos} catalogue) to estimate masses,
radii, gravities, and effective temperatures for 17,219 stars that
appear   within the range of masses, radii, and gravities assessed in
this study out of  22,982 stars included in the {\it Hipparcos}
catalogue within the  sphere of 100 pc radius from the Sun.  Solar
metallicity ($-0.4 <$ [Fe/H] $< +0.4$) has been assumed. Although the
tail of the metallicity distribution of stars in the solar
neighbourhood reaches [Fe/H] $\simeq -1$,  most of the stars are within
the selected interval (Rocha-Pinto \& Maciel 1996).

\begin{figure}
      \vspace{0cm}
	\hspace{0cm}\psfig{figure=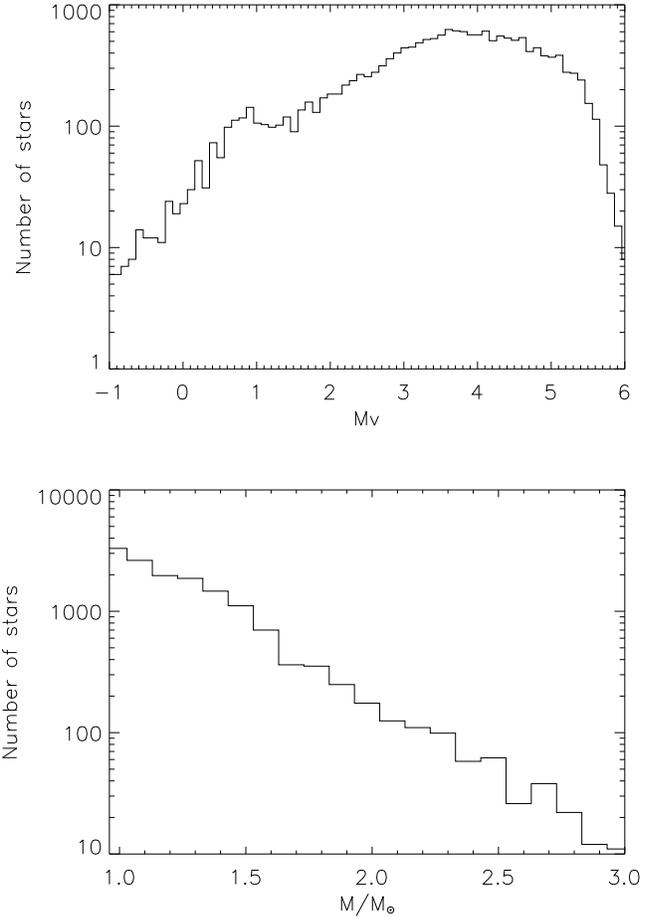,width=8.5cm}
      \caption[]{Luminosity function (upper panel) and mass function (lower panel) for the sample studied. 
	}
      \label{Fig6}
      \vspace{0cm}
\end{figure}

An overview of a few of the possible uses of these data is given in the
next section. Table 1, available only in electronic format,  includes
the following information: {\it Hipparcos} \#, V, $\pi$, $\sigma(\pi)$,
$M_v$, $\sigma(M_v)$, B$-$V, $\log g$, $\sigma(\log g)$, M/M$_{\odot}$,
$\sigma({\rm M/M}_{\odot})$, $\log (R/R_{\odot})$, $\sigma(\log R/R_{\odot})$, $BC$, $\sigma(BC)$,  $\log T_{\rm eff}$, and $\sigma(\log
T_{\rm eff})$. Fig. \ref{Fig6} shows the derived luminosity and actual
mass functions for the sample.  These figures have been derived using
the data in Table 1, and therefore are restricted to the range $0.8 \le
{\rm M}$/${\rm M}_{\odot}  \le 22$, although the statistics are poor
for the more massive stars and we have further restricted the plots.
 Although it is beyond the scope of this paper, the study of these
distribution functions will likely shed light on the discussion about
the universality of the initial mass function, and the star formation
history of the Galactic disc. The use of a finer grid for the
interpolation in stellar age and metallicity may provide additional
information.

\section{Discussion of the results}

The comparison between the stellar masses, radii and effective
temperatures
 for detached eclipsing binaries and those analyzed by the IRFM with
 the values retrieved from the position of the stars in the
colour-magnitude diagram and interpolation in the 
isochrones of Bertelli et al. (1994), showed that  the latter
provides a method to estimate those fundamental parameters with high
accuracy for most purposes. Stellar radii can be predicted to $\sim
6$\%, masses to $\sim 8$\%, and effective temperatures to $\sim 2$\% 
for the ranges in these parameters listed in \S4.

\begin{figure*}
      \vspace{0cm}
	\hspace{0cm}\psfig{figure=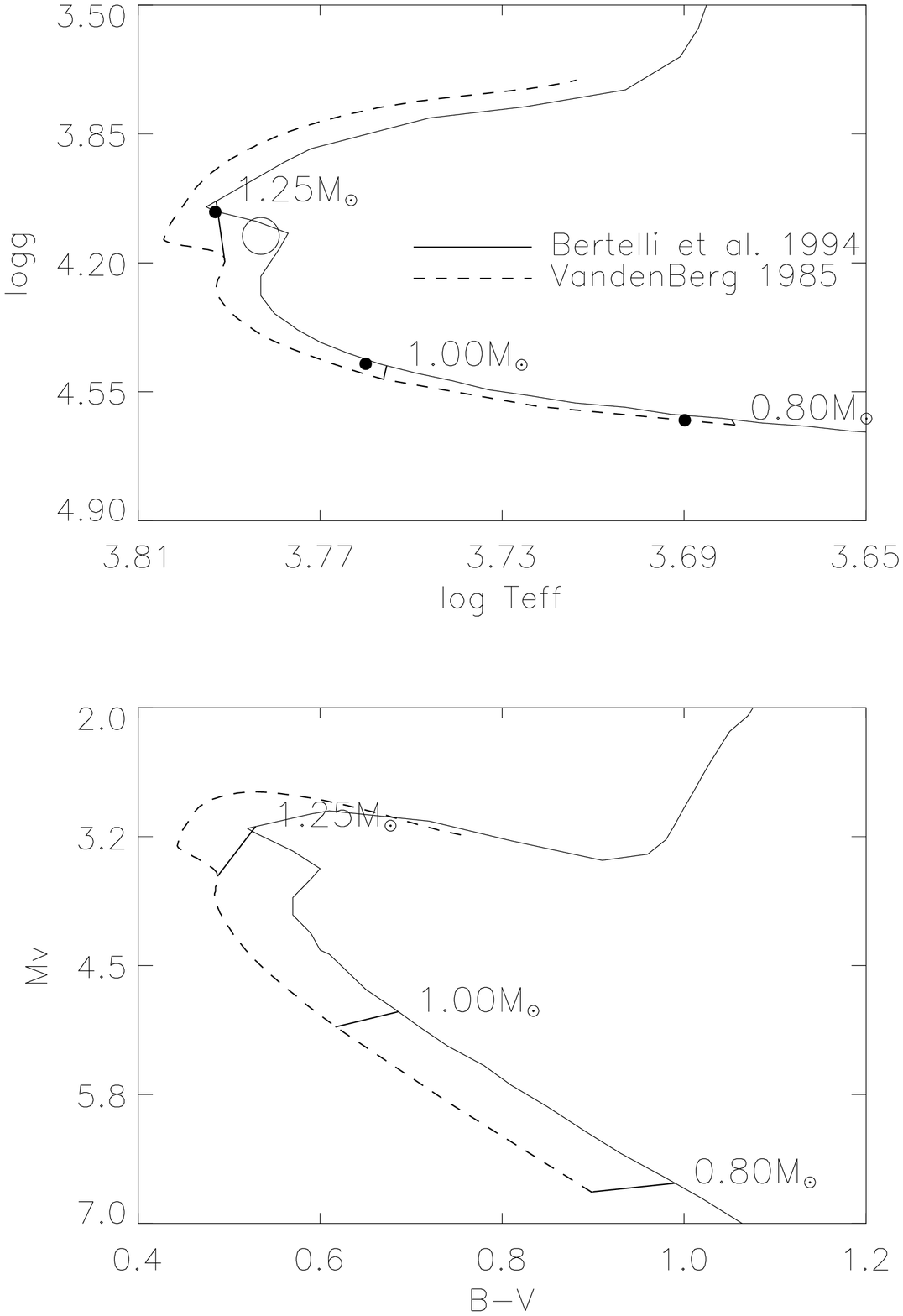,width=8.5cm}
	\hspace{0cm}\psfig{figure=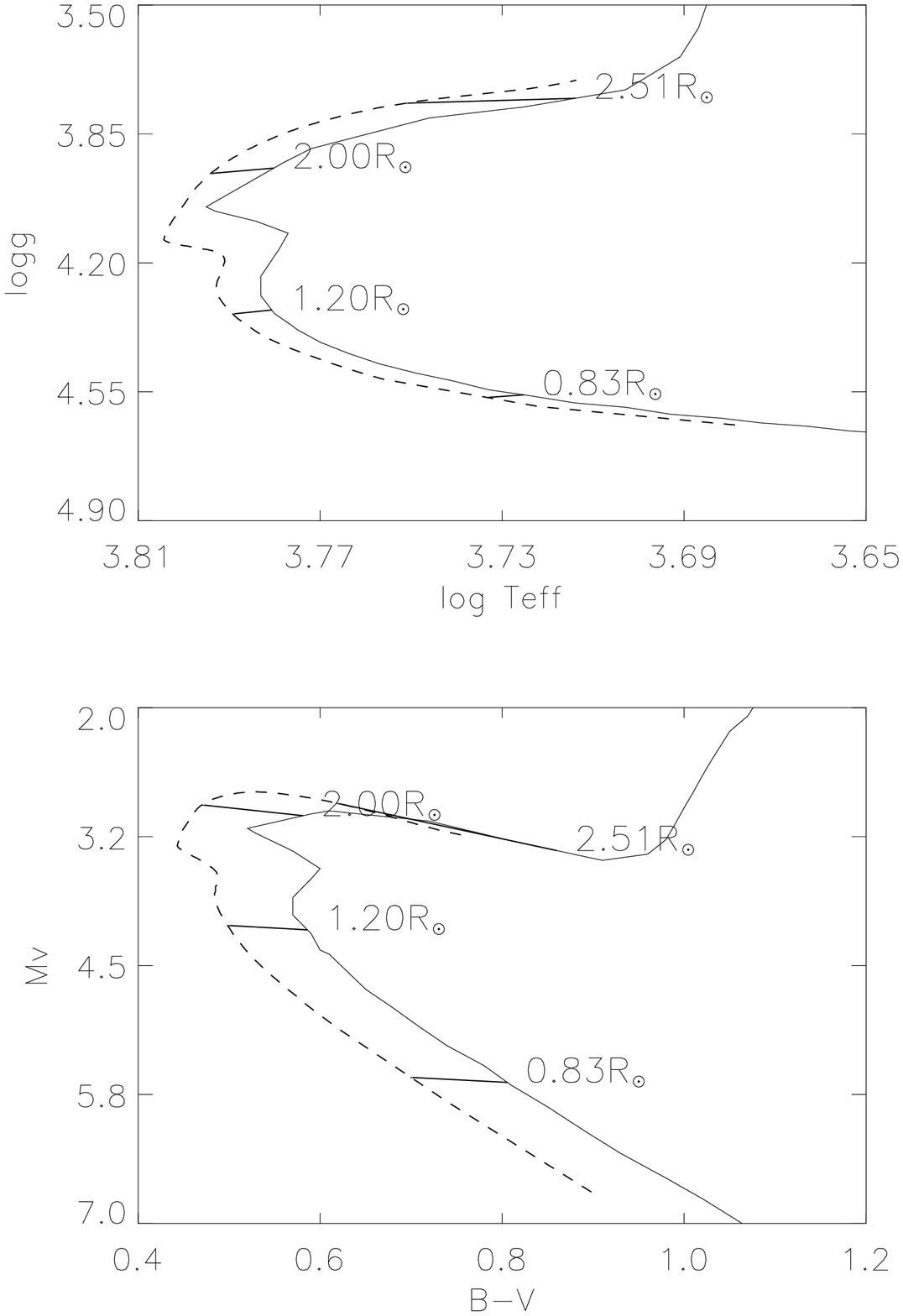,width=8.5cm}
      \caption[]{Differences between the isochrones of Bertelli et al. (1994; 
solid line) and VandenBerg (1985; dashed line) for solar metallicity and an 
age of 4 Gyrs. Solid segments join the positions of stars with masses of
0.8, 1.0, and 1.25 M$_{\odot}$ (left panels) or radii of 0.83, 1.2, 2.0, 2.51
R$_{\odot}$ (right panels) in both sets of isochrones. 
The location of stars with the same masses in the theoretical ($\log g-T_{\rm eff}$) plane for the (non-overshooting) calculations of Schaller et al. (1992) 
are marked with filled circles, including also the case for a 1.25 M$_{\odot}$
model with overshooting (open circle).
	}
      \label{Figl}
      \vspace{0cm}
\end{figure*}

However, the longest episode in  stellar evolution, the
main-sequence, imposes a bias in the samples. Giants and super-giants
are not  common in the analysed samples with high-quality data, and
hence our conclusions cannot be safely extended to very low gravities.
New observational efforts are in progress (see, e.g. Bennett, Brown \&
Yang 1998) and are likely to improve the situation in the future.  The
same or very similar arguments apply to the metal content of the stars.
Our sample is biased towards solar-like metallicities and from the
material studied here it is not possible to derive any conclusion for
the lower-metallicity domain.  The addition of more constraints, such
as prior (from spectroscopy or photometry) knowledge of the stellar
metal abundance, and a more detailed analysis, weighting the different
possible stellar models with care, depending on the speed with which
they cross that part of the colour-magnitude  diagram,  should likely
improve the results and possibly extend its applicability to metal-poor
stars.

Rotation changes the position of the stars in the colour-magnitude diagram
(Maeder \& Peytremann 1970). Typically, the position of
 a star rotating at $\sim 200$  km s$^{-1}$ will change by 0.1--0.3 mag in
$M_v$ and 200--250 K in $T_{\rm eff}$. 
Therefore, for high rotational velocities,  has to be taken into  
account before  applying the procedure described here. 

It is possible to  get better B--V colour estimates than those compiled
in the {\it Hipparcos} catalogue. The calibration of Harmanec (1998)
makes it possible to accurately estimate the V and B colours from the
B$-$V and U$-$B indices together with the {\it Hipparcos} $H_p$
magnitudes. Besides, the B--V colours listed in the {\it Hipparcos}
catalogue can be refined by combining them with Earth-based
measurements, as proved by Clementini et al. (1999).

We have selected the set of isochrones derived from the evolutionary
calculations of Bertelli et al. (1994) because they are one of the most
homogeneous and comprehensive among those publicly available in
electronic format. It is of interest to check whether the use of
alternative models would lead to the same conclusions. Considering the
particular case of stars between 0.8 and 1.25 M$_{\odot}$ of 4 Gyr. age, we
can compare the calculations of VandenBerg (1985) with those of
Bertelli et al. (1994).  Figure \ref{Figl} displays such a comparison
in both, the theoretical ($\log g - T{\rm eff}$) and observational
($M_v -$ B--V) planes. The points in the isochrones corresponding to
equal masses (left-side panels; 0.8, 1.0, 1.25 M$_{\odot}$) or radii
(right-hand panels; 0.83, 1.20, 2.0, 2.51 R$_{\odot}$) have been linked
by solid segments.  Differences in the theoretical plane may be the
effect of one or more of the different ingredients in the calculations,
such as  convection treatment or radiative opacities (Los Alamos
 Opacity Library vs. OPAL), as well as slightly  different assumed
metallicities (Z=0.0169 vs. 0.02), or mass fraction of helium (Y=0.25
vs. 0.28). The agreement is  better for the dwarfs with  lower
effective temperatures and gets poorer for lower gravities. 
However, the discrepancies in the observational plane  become more
significant and systematic, and could induce important differences in
the results. Even though there are details such as discordant
assumptions for the Sun's bolometric correction ($-0.12$ vs.  $-0.08$),
the different model atmospheres employed in the translation from the
theoretical magnitudes must play a major role. The more recent model
atmospheres (Kurucz 1992) employed by Bertelli et al. perform
adequately, as suggested by the conclusions in \S3.


The positions in the $\log g - T_{\rm eff}$ plane predicted by the
 calculations of Schaller et al. (1992) without overshooting, shown
with filled circles in the left-side upper panel of Fig. \ref{Figl} for
0.8, 1.0, and 1.25 M$_{\odot}$, do not exactly overlap neither with the
predictions derived from Bertelli et al.'s calculations, nor with those
of VandenBerg (1985), yet even so they include the same radiative
opacities (LAOL; Huebner et al. 1977) and assume the same value for the
mixing-length $\alpha$ as VandenBerg's models. Again, slightly
different values for Z$_{\odot}$, Y$_{\odot}$, and  details in the
treatment of envelope convection must be responsible. The open circle
correspond to a 1.25 M$_{\odot}$ model with overshooting.

Balona (1994) made use of the calibration of Balona \& Shobbrook (1984) 
to estimate absolute magnitudes from the synthetic Str\"omgren indices
computed by Lester et al. (1986) based on Kurucz (1979) model
atmospheres. He derived effective temperatures,
gravities, and  luminosities from the model atmospheres, and used the
$T_{\rm eff}$s and luminosities to interpolate in the 
models of stellar evolution by Claret \& Gimenez (1992) 
and Schaller et al. (1992) and then estimate {\it evolutionary} gravities.  The
comparison showed that the {\it evolutionary} gravities were larger than
 the gravities from the model atmospheres for stars with $\log g < 4$, and
the discrepancy was increasing towards lower gravities. Besides, the
situation appeared to be reversed for stars with $\log g > 4$. The 
evolutionary calculations used here tend to underpredict the stellar masses
for low-gravity stars. However, this effect is much smaller and in opposite
sense to the huge differences found by Balona (1994) that were as large 
as 0.5 dex for $\log g$ ({\it evolutionary}) $\sim 3.5$. The explanation 
is still unclear, but we note that the comparison of the gravities from
LTE (model-atmospheres) spectroscopic analysis 
(iron ionization equilibrium)  with those obtained combining 
{\it Hipparcos} parallaxes and evolutionary  models  seem to agree 
reasonably well (Allende Prieto et al. 1999; see also Fuhrmann 1998).

Gravitational redshifts are proportional to the mass-radius ratio and
systematically affect measurements of stellar radial velocities
(Dravins et al. 1999). Photospheric spectral lines are  shifted by
roughly 600 m s$^{-1}$ for a star like the Sun, and by more than 1000 m
s$^{-1}$ for more massive stars in the main sequence. The empirically
determined errors for masses and radii derived from the position of a
star in the colour-magnitude diagram make it possible to estimate its 
gravitational redshift with an uncertainty of the order of 100 m s$^{-1}$.  

A precise knowledge of the
stellar radius and the distance to the star combine together to
translate the measured stellar flux to the flux at the star's surface,
a quantity that can be compared with  synthetic spectra from model
atmospheres. Spectrophotometry  is already available for the large
sample of stars observed by IUE, HST, and many other missions, and
their investigation will likely provide valuable information on the
stellar content of the Galactic disc in the solar neighbourhood and
beyond.

\begin{acknowledgements}

This work has been partially funded by the  NSF (grant AST961814)
  and the Robert A. Welch Foundation of Houston, Texas. We thank Mario
M. Hern\'andez for fruitful discussions about rapidly rotating stars.
We are as well indebted to the the referee, Gianpaolo Bertelli, for
 helpful comments that improved the paper. We have made use of data
from the {\it Hipparcos} astrometric mission of the ESA, the NASA ADS,
and  SIMBAD.

\end{acknowledgements}


\begin{thebibliography}{}


\bibitem[1999]{allende} Allende Prieto, C., Garc\'{\i}a L\'opez, R. J., Lambert, D. L.,  Gustafsson, B., 1999, ApJ, in press

\bibitem[1982]{landolt-bornstein} Aller, L. H., Appenzeller, I., Baschek, B., Duerbeck, H. W., Herczeg, T., Lamla, E., Meyer-Hofmeister, E., Schmidt-Kaler, T., Scholz, M., Seggewiss, W., Seitter, W. C., Weidemann, V., 1982, 
Landolt-B\"ornstein: Numerical Data and Functional Relationships in Science and Technology - New Series, Group 6, Vol. 2, Stars and Star Clusters (New York: Springer-Verlag)

\bibitem[1991]{andersen} Andersen, J., 1991, A\&AR 3, 91


\bibitem[1994]{balona} Balona L. A., 1994, MNRAS 268, 119

\bibitem[Balona \& Shobbrook 1984]{1984MNRAS.211..375B} Balona, L. A., 
Shobbrook, R. R., 1984, MNRAS 211, 375 


\bibitem[Bennett Harper Brown \& Hummel 1996]{1996ApJ...471..454B} Bennett, 
P. D., Harper, G. M., Brown, A.,  Hummel, C. A., 1996, ApJ 471, 454 

\bibitem[Bennett Brown \& Yang 1998]{1998psrv.confE..61B} Bennett, P. D., 
Brown, A.,  Yang, S.,  1998, IAU Colloq. 170: Precise Stellar Radial 
Velocities, E61 


\bibitem[Bertelli et al. 1994]{1994A&AS..106..275B} Bertelli, G., Bressan, 
A., Chiosi, C., Fagotto, F.,  Nasi, E., 1994, A\&AS 106, 275 

\bibitem[1994]{blg} Blackwell, D. E., Lynas-Gray, A. E., 1994, A\&A 282, 899

\bibitem[Claret \& Gimenez 1992]{1992A&AS...96..255C} Claret, A.,  
Gimenez, A., 1992, A\&AS 96, 255 


\bibitem[Clementini Gratton Carretta \& Sneden 1999]{1999MNRAS.302...22C} 
Clementini, G., Gratton, R. G., Carretta, E.,  Sneden, C., 1999, MNRAS 
302, 22

\bibitem[]{}Dravins, D., Gullberg, D., Lindegren, L., Madsen, S. 1999, In: Precise Stellar Radial Velocities,  J. B. Hearnshaw  and C.D. Scarfe, eds., ASP Conference Series, in press

\bibitem[Goriely \& Clerbaux 1999]{1999A&A...346..798G} Goriely, S.,  
Clerbaux, B., 1999, A\&A 346, 798 

\bibitem[Harmanec 1998]{1998A&A...335..173H} Harmanec,  P., 1998, A\&A 335, 
173

\bibitem[]{} Huebner, W. F., Merts, A. L., Magee, N. H., Argo, M. F.,  1977, Los Alamos Sci. Lab. Rept., LA-6760-M

\bibitem[Iglesias Rogers \& Wilson 1992]{1992ApJ...397..717I} Iglesias, C. 
A., Rogers, F.  J.,  Wilson, B. G., 1992, ApJ 397, 717 

\bibitem[Kurucz 1979]{1979Kurucz} Kurucz, R. L., 1979, ApJS 40, 1

\bibitem[Kurucz 1992]{1992Kurucz} Kurucz, R. L., 1992, private communication

\bibitem[1970]{1970MaederPeytremann} Maeder, A.,  Peytremann, E., 1970,
 A\&A 7, 120

\bibitem[Perryman et al. 1995]{1995A&A...304...69P} Perryman, M. A. C., Lindegren, L., Kovalevsky, J., Turon, C., 
H$\o$g, E., Grenon, M., Schrijver, H., Bernacca, P. L., Creze, M., Donati, 
F., Evans, D. W., Falin, J. L., Froeschle, M., Gomez, A.,  Grewing, M., Van Leeuwen,  F., Van der Marel, J., Mignard, F., Murray, C. A., Penston, M. J., Petersen, C. S., Le Poole, R. S., Walter, H.,  1995, A\&A 304, 69  


\bibitem[1997]{pols} Pols, O. R., Tout, C. A., Schr\"oder, K.-P., Eggleton, P. P., Manners, J., 1997, MNRAS 289, 869


\bibitem[1980]{popper} Popper, D. M., 1980, ARA\&A 18, 115

\bibitem[Popper \& McAlister 1987]{1987AJ.....94..700P} Popper, D. M., 
McAlister, H. A., 1987, AJ 94, 700 


\bibitem[Richichi Ragland Stecklum \& Leinert 1998]{1998A&A...338..527R} 
Richichi, A., Ragland, S., Stecklum, B.,  Leinert, C.,  1998, A\&A 338, 527

\bibitem[Rocha-Pinto \& Maciel 1996]{1998MNRAS.298..332R} Rocha-Pinto, H.  J.,   Maciel, W. J., 1996, MNRAS 279, 447

\bibitem[Rocha-Pinto \& Maciel 1998]{1998MNRAS.298..332R} Rocha-Pinto, H.  J.,   Maciel, W. J., 1998, MNRAS 298, 332


\bibitem[Rogers \& Iglesias 1992]{1992ApJ...401..361R} Rogers, F. J.,  
Iglesias, C. A., 1992, ApJ 401, 361 

\bibitem[Schaller Schaerer Meynet \& Maeder 1992]{1992A&AS...96..269S} 
Schaller, G., Schaerer, D., Meynet, G.,  Maeder, A., 1992, A\&AS 96, 269 


\bibitem[1997]{schroder} Schr\"oder, K.-P., Pols, O. R., Eggleton, P. P., 1997, MNRAS 285, 696

\bibitem[1985]{vandenberg} VandenBerg, D. A., 1985, ApJS 58, 711

\end{thebibliography}
\end{document}